\documentclass[%
reprint,
 amsmath,amssymb,
aps,
]{revtex4-2}

\usepackage{graphicx}
\usepackage{dcolumn}
\usepackage{bm}
\usepackage[normalem]{ulem}
\usepackage{hyperref}

\usepackage{physics}
\usepackage{xcolor}

\begin{document}


\title{Black holes in a dense infinite medium:\\
a toy-model regularizing the Schwarzschild metric}

\author{Aurélien Barrau}
\author{Killian Martineau}
\author{Hanane Zelgoum}

\affiliation{
 Laboratoire de Physique Subatomique et de Cosmologie, Univ. Grenoble Alpes, CNRS/IN2P3\\ 53 avenue des Martyrs, 38026 Grenoble Cedex, France
}

\date{\today}

\begin{abstract}
We revisit the dynamics of a black hole accreting energy from a surrounding homogeneous and infinite space. We argue for a simple heuristic modification of the Schwarzschild approximation when the density of the medium is not negligible anymore. The resulting behavior is drastically modified: the mass divergence at finite time is cured and the thermodynamical properties are deeply changed. Some potential consequences for quantum gravity and bouncing models are also pointed out. Those conclusions being mostly obtained from a Newtonian approach, they only aim at guiding toward a more rigorous treatment. Still, interestingly, the behavior is far more convincing that the one usually obtained.
\end{abstract}

\maketitle

\section{Introduction}

It is well known that the Schwarzschild metric leads to a curious mass divergence at finite time for a black hole immersed in a thermal bath. This has first been noticed in \cite{Novikov2} (available in English in \cite{Novikov}). A large number of studies were devoted to the so-called self-similar solution, taking into account different equations of state, investigating the existence of a Friedmann or quasi-Friedmann asymptotic behavior, and considering the separate universe issue \cite{Carr:1974nx,Carr:1975qj,bick,carr,Maeda:2002vq,lin,hacyan,Madsen:1988ph,Harada:2004pe,Harada:2006dv,Maeda:2007tk,Kyo:2008qi,Carr:2010wk,Kopp:2010sh,Carr:2014pga}. The entire picture was recently reconsidered in \cite{Barrau:2022bfg}.\\

This ``pathological" evolution does not come as a full surprise as the Schwarzschild solution is not anymore an appropriate approximation when the density of the surrounding medium is not negligible. The correct solution however remains unknown although many attempts and incomplete results are available (see, {\it e.g.}, \cite{Husain:1995bf,Kiselev:2002dx,nielsen,Babichev:2004yx,Nielsen:2005af,Sultana:2005tp,Faraoni:2007es,Gao:2008jv,Faraoni:2008tx,Liu:2009ts,Faraoni:2009uy,Vanzo:2011wq,Abdalla:2013ara,Guariento:2015cxa,Thakurta,Heydarzade:2016zof,Babichev:2018ubo,Faraoni:2018xwo,Boehm:2020jwd,Kobakhidze:2021rsh,Harada:2021xze,Boehm:2021kzq,Xavier:2021chn} and references therein). This is why any ``guide" is potentially useful, hence this work.\\

In this article, we do not aim at giving a definitive general relativistic answer to the question of the fate and structure of a black hole in a homogeneous and infinite medium. We simply suggest a Newtonian toy-model modification to the crude Schwarzschild-like picture (which is, of course, not an exact solution outside the vacuum), taking into account, in the simplest possible way, the specificity of the considered situation. Our results should therefore be considered with great care, mostly as ``hints" for a more accurate treatment. Still, heuristically, this approach works in the vacuum case, which gives hope that it us not meaningless.
Quite surprisingly, the resulting picture is rich and plausible. The goal is neither phenomenological nor mathematical: it is only to give some hints for constructing a possibly improved picture. We explicitly calculate the dynamics, the temperature and the  evolution of a binary system of black holes. We also speculate about possible quantum gravity consequences and links with bouncing models. It is shown that all pathological behaviors are cured. Beyond the question of the mass evolution in an infinite homogeneous medium, we believe that this no-nonsense model may help finding a more accurate and rigorous solution. 
\section{The usual picture}

Let us begin by recalling the usual elementary picture. The Schwarzschild metric is assumed to describe correctly (although approximately outside vacuum) a black hole in a thermal bath. The growth rate of the black hole mass is proportional to its area:
\begin{equation}
    \frac{dM}{dt}=\lambda R_H^2 \frac{\rho}{c} ,
    \label{dmdt1}
\end{equation}
where $R_H$ is the horizon radius, here assumed to be the Schwarzschild radius, $R_S=2GM/c^2$, $\rho$ is the energy density of the surrounding medium, and $\lambda$ is a dimensionless constant. When taking into account the fact that the effective cross section of the black hole is slightly larger than its area due to the bending of space, and integrating out over the appropriate solid angle, one gets $\lambda=27$. This leads to the mass evolution: 
\begin{equation}
    \frac{dM}{dt}=\beta \frac{G^2}{c^5}\rho M^2,
\label{dmdt}
\end{equation}
where $\beta$ is also a dimensionless constant (equals to 108). This leads to:
\begin{equation}
    M(t)=\frac{1}{M_{init}^{-1}-\beta\frac{G^2\rho}{c^5} t},
    \label{usualevol}
\end{equation}
where $M_{init}$ is the mass of the black hole at $t=0$. This implies a divergence at {\it finite} time $t_{div}=c^5/(\beta M_{init}G^2\rho)$. No mathematical mystery here: as soon as $\gamma >1$, any differential equation of the form $df/dt=f^{\gamma}$ trivially diverges at finite time. The situation however remains physically puzzling: does it really make sense? Does the black hole fills the entire space at finite time? In particular, this model implies that at some point the black hole inevitably become {\it less} dense that the surrounding medium. Is this meaningful? Is it, at least, a correct approximation? The aim of this work it to provide a tentative answer.

\section{Back to basics}

Let us begin with an extremely simple question: what is the gravitational field at the surface of an empty hole of radius $R_H$ in an infinite homogeneous medium in a purely Newtonian approach? We proceed very slowly as the result might seem surprising. \\

\begin{figure}
    \centering
    \includegraphics[width=.9\linewidth]{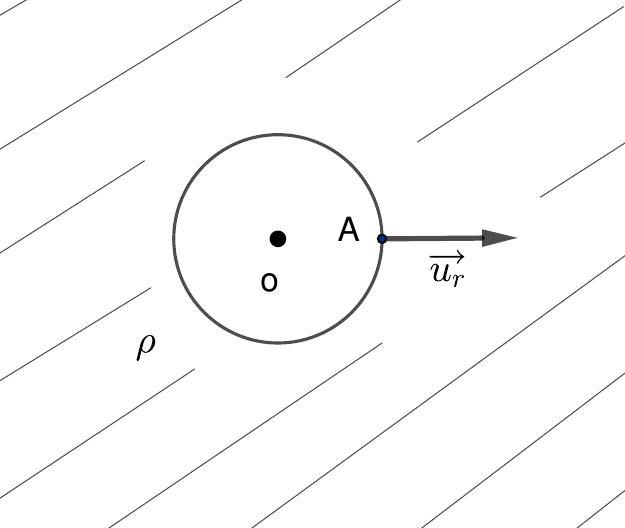}
    \caption{Sketch of the Gauss approach to the question: the gravitational field at the surface of the hole vanishes.}
    \label{fig1}
\end{figure}

The Gauss theorem leads to a clear answer (see Fig. \ref{fig1}). The problem being spherically symmetric from $O$, the center of the hole, the gravitational field $\vec{\mathcal{G}}$ depends only upon the radial coordinate $r$. In addition, planes containing points $O$ and $A$ are symmetry planes, hence they contain $\vec{\mathcal{G}}$. The field therefore writes $\vec{\mathcal{G}}=\mathcal{G}(r)\vec{u_r}$. So, the Gauss theorem can be easily applied, leading to $4\pi R_H^2\mathcal{G}(R_H)=-4\pi GM_{int}$ at the surface of the hole, where $M_{int}$ is the mass inside the hole. As $M_{int}=0$, this immediately leads to $\mathcal{G}(R_H)=0$. The gravitational field at the surface of the hole vanishes. Clean and simple.\\

\begin{figure}
    \centering
    \includegraphics[width=.9\linewidth]{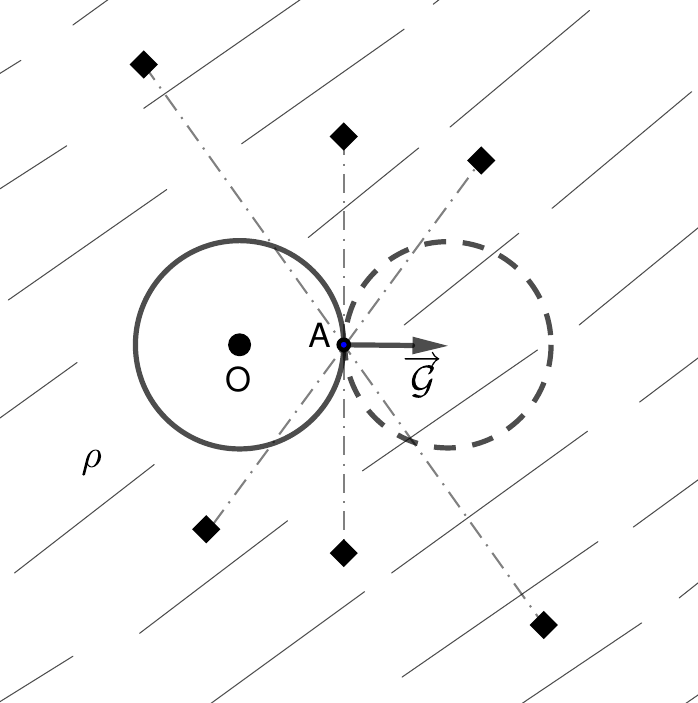}
    \caption{Sketch of the alternative approach to the same question: he gravitational field at the surface of the hole does {\it not} vanish.}
    \label{fig2}
\end{figure}

Let us now take another robust view on the very same situation, illustrated by Fig. \ref{fig2}. Any volume element in space generates a field at the point $A$ which is exactly compensated by the symmetrical (with respect to $A$) volume element. This is true for the entire space but for points within the sphere symmetrical to the hole, represented by a dotted line on Fig. \ref{fig2}. Therefore the point $A$ feels a non-vanishing gravitational field simply given by the ``non-compensated" sources: $\mathcal{G}=G(\frac{4}{3}\pi R_H^3\rho/c^2)/R_H^2=\frac{4}{3}\pi G R_H\rho/c^2$.\\

At this point, it seems that we have two clear and elementary ways of thinking leading to incompatible results. Other version of the ``paradox" are given in the appendix but the argument presented here is enough to proceed.\\

In a sense, both conclusions are correct. There is no trivial mistake. The only difference is the way infinity is approached. In the first case, the entire argument is centered on the hole and the result is indeed correct whatever the cutoff $R_{max}$ possibly imposed on the surrounding medium, as long as the latter is a sphere centered on the hole. In the second case, the argument also holds whatever $R_{max}$, but as long it is centered on the observer located at point $A$. This simple remark is also the key solution to the examples given in the appendix. The strange ``discontinuity" which appears, in some circumstances, in the evolution of the field is entirely due to symmetries being applied from another point. Legitimately so (when space is fully homogeneous, for example, there is nothing wrong in choosing an arbitrary symmetry center)! 
This means that in an infinite homogeneous medium, no conclusion can be reached without picking up a privileged point. In the case of the hole considered here, there are 2 obvious candidates: the center of the hole, which is the ``symmetry center of the Universe", or the observer (located on the edge of the hole).\\

This digression aimed at making the following point. The usual Schwarzschild metric description to the growing black hole (initiated by Zel’dovich and recently exhaustively studied in \cite{Barrau:2022bfg}) in an infinite medium is fundamentally rooted on the ``Gauss theorem" view. Even when the black hole is an under-density (let us now assume there is a non-vanishing mass $M$ located in $0$), the Gauss theorem tells us that the gravitational field in indeed attractive at the edge, directed toward the center. Even more: it tells us that the entire exterior universe has no influence at all and that the position of the (Newtonian) horizon is therefore exactly given by the Schwarzschild solution $R_S=2GM/c^2$. 

This is however slightly disturbing: as the black hole grows its density decreases and at some point it inevitably become less dense than the surrounding medium. Can the observer situated at $R_S$ from the center really be attracted by the under-density? When the black hole is huge its density is tiny: is a particle on the edge attracted by a nearly empty hole -- that is by the precise place in the universe where there is {\it less} attracting mass? Otherwise stated: does the horizon have a chance to be really located -- even approximately -- at the Schwarzschild radius?\\

The full resolution of the problem in general relativity is notably difficult and way beyond the scope of this work. At the level of a toy-model, we advocate the simple idea that the field felt by a test particle is to be evaluated from the viewpoint of the particle itself. Basically, this implies that if the horizon is defined as the spacelike surface where the escape velocity reaches the speed of light -- accounting for the local gravitational potential -- then, in an infinite homogeneous medium, it does not coincide with the Schwarzschild radius.
Going back to the previous discussion, this means that we choose the second point of view.

Should a regulator be included in the form of a maximum distance -- a kind of Newtonian equivalent of the Hubble radius --, it should be centered on the observer.
In the following, we therefore define the horizon as the surface where the escape velocity is equal the speed of light, the field being evaluated {\it from the point of view of the considered location}. This is arguably the most reasonable assumption. As the simple argument given above shows, the edge of an under-dense ball in an infinite medium can obviously {\it not} be a horizon from the view-point of an observer at the edge: just the other way around, she feels a (classical) force directed outside of the ball. However, as she approaches point $O$, she  will inevitably reach a horizon. Where is this horizon located?

\section{The model}

The usual way to define the Newtonian horizon is to basically write $E_p=E_c$, with $E_p$ the attractive potential energy due to the mass $M$, and $E_c$ the (classical) kinetic energy at the speed of light. This immediately leads (heuristically) to the Schwarzschild radius $R_S=2GM/c^2$. The argument can be straightforwardly adapted to the case of interest here, following the second prescription of the previous section. We insist on recalling that the approach is, on purpose, naive and simple. The new argument then reads $E_p=E_p'+E_c$, where $E_p'$ is the potential energy associated with the attraction exerted on the point $A$ (in Fig. \ref{fig2}) by the sphere on the right side (that is by the entire universe as all other points, but within the black hole, do not contribute), and $E_p$ is the  one associated with the mass M now assumed to be in $O$ (that is the black hole). Otherwise stated, the idea is simply that a horizon is reached once the attraction due to M overwhelms not only the kinetic energy of the escaping particle (usual requirement) but also the effect of the environment.\\

This translates into: 
\begin{equation}
    G\frac{M}{R_H}=\frac{4}{3}\pi R_H^3 \left( \frac{\rho} {c^2} \right) \frac{G}{R_H}+\frac{1}{2}c^2.
    \label{eq1}
\end{equation}
When $\rho =0$, one hopefully immediately recovers the Schwarzschild solution
\begin{equation}
R_H^{vac}=R_S=\frac{2GM}{c^2}.
\end{equation}
In the other limit, when the second term is much larger than the third one -- that is when the energy density of the medium plays a dominant role --, the horizon is given by 
\begin{equation}
    R_H^{dens}=\left( \frac{3Mc^2}{4\pi \rho} \right)^{1/3}.
    \label{secondH}
\end{equation}
Most of the following work focuses on this case, which is obviously the interesting one.
\\

Several comments are in order. First, it should be noticed that, in this latter case, the horizon is much closer to the center of the black hole than the Schwarzschild horizon. This is expected. Once again, we emphasize that from the point of view of the observer at the horizon, all points in space do compensate each other by opposite gravitational fields except for those located in the black hole and in the symmetrical sphere. The latter pulls away from $M$ and it is therefore natural that one has to get closer to $M$ to reach the real horizon. 

Second, Eq. (\ref{secondH}) depends only upon the speed of light because $\rho$ is an energy density. Should we have written everything as a function of the mass density of the medium, the speed of light would have disappeared. It seems weird: should we have required the liberation speed to be $5c$ or $10c$, we would have defined nearly the same horizon position in this regime -- which is obviously not true for the Schwarzschild horizon. Although disturbing at first sight, this is once again expected: in this case the potential energies involved are so much larger than the kinetic energy that adding the latter does not change the picture. This basically means that we are in a region where the gravitational field varies so rapidly with the distance that, as soon as the potential energy of $M$ dominates, nearly no speed can counter-balance the attraction. 

Third, in this regime, the horizon radius scales as $M^{1/3}$ -- as usual matter -- and not as $M$. This is the big difference with the usual view. As a consequence, the density of the black hole now scales as $M^0$ (and as $\rho ^1$) whereas, for the Schwarzschild black hole, it scales as $M^{-2}$ (and as $\rho^0$), $\rho$ being the density of the surrounding medium. It should also be pointed out that the surface gravity now increases with $M$, as $M^{1/3}$, whereas it decreases as $M^{-1}$ in the Schwarzschild case.\\

In the intermediate regime, the general solution $R_H$ of the cubic equation 
\begin{equation}
R_H^3+\frac{3R_Hc^4}{8\pi G\rho}-\frac{3Mc^2}{4\pi\rho} = 0   
\end{equation}

lies between $R_H^{vac}$ and $R_H^{dens}$. It is given by 
\begin{equation}
    R_H = 2 \sqrt{ \frac{c^4}{8 \pi G \rho} } 
\sinh\left( \frac{1}{3} \sinh^{-1} \left( \frac{3 G M}{c^2} \sqrt{ \frac{8 \pi G \rho}{c^4} } \right) \right).
\label{fulleq}
\end{equation}

\begin{figure}
    \centering
    \includegraphics[width=.9\linewidth]{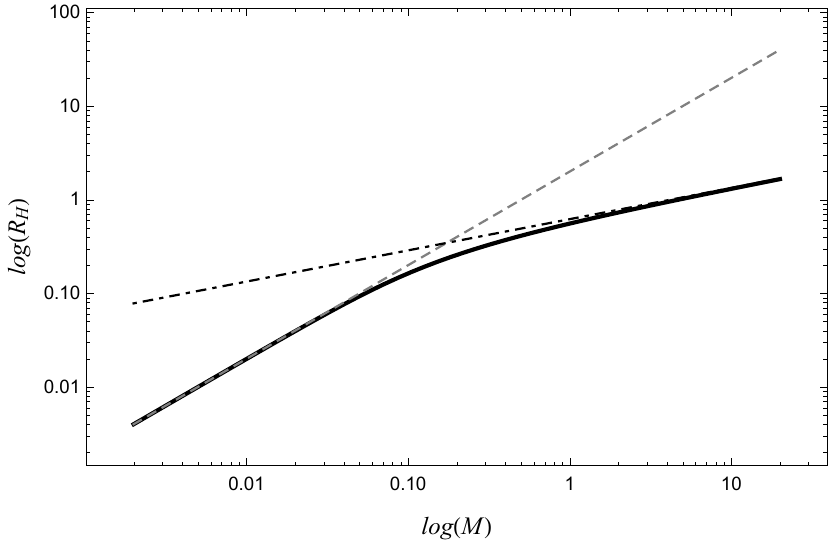}
    \caption{Horizon radius as a function of the mass, in double logarithmic scale and in Planck units -- setting $\rho=1$. The exact solution is in solid black, the low mass approximation is in gray dashed, and the large mass approximation is in black dot-dashed.}
    \label{fig3}
\end{figure}

The shape of $R_H(M)$ is displayed on Fig. \ref{fig3}. 
The transition between the $R_H\propto M$ and the $R_H\propto M^{1/3}$ regimes roughly happens when the second and the third terms in Eq. (\ref{eq1}) are of the same order of magnitude. This corresponds to a critical radius 
\begin{equation}
    R_H^{c}\sim \left( \frac{3c^4}{8\pi G \rho} \right)^{1/2},
\end{equation}
associated with a critical mass 
\begin{equation}
    M_c\sim\frac{c^2}{G}\left( \frac{3c^4}{8\pi G \rho} \right)^{1/2}.
    \label{M_c}
\end{equation}

Although the entire construction is no more than a toy-model, the global image is surprisingly consistent. Interestingly -- and this is quite obvious when taking into account the way the horizon is here defined  -- the density of the black {\it cannot} become less than the one of the surrounding medium. It asymptotically approaches the latter for very large masses, quite satisfactorily. This is to be contrasted with the use of the Schwarzschild solution for which the black hole density becomes less than $\rho$ for $M>(3c^8/(32\pi G^3\rho)$.\\

It should also be pointed out that the horizon radius, in the large mass limit given by Eq. (\ref{secondH}) can, in principle, {\it decrease} with time. Of course, as it should, the mass trapped in the black hole can only increase. However, if the surrounding medium was such that $\rho$ were to increase with time -- for example in a contracting universe --, $R_H$ would decrease, which is very consistent with the approach followed in this work: the denser the medium, the closer to the center of the black hole the horizon has to be. This might appear as problematic from the viewpoint of \cite{Hawking:1971tu} but we believe that the situation is actually more akin to that of a Schwarzschild–de Sitter black hole, whose horizon radius would also decrease -- in these coordinates -- if the cosmological constant were to decrease.

\section{Growth of the regularized black hole}

Let us come back to the initial point: how does such a black hole grows inside a radiation bath of energy density $\rho$? Equation (\ref{dmdt1}) still holds but $R_H$ is now given by Eq. (\ref{fulleq}).\\

In the small mass limit, the usual behavior given by Eq. (\ref{usualevol}) is obviously recovered, that is
\begin{equation}
    M(t)=\left( M_{init}^{-1}-\beta\frac{G^2\rho}{c^5} t \right)^{-1}.
\end{equation}
This regime roughly holds (assuming $M_{init}<M_c$) until the critical time
\begin{equation}
    t_c\sim\frac{c^5}{\beta G^2 \rho} ( M_{init}^{-1}-M_c^{-1} ),
\end{equation}
such that $M$ reaches $M_c$, which is given by Eq. (\ref{M_c}). As $t_c$ is here evaluated using the first behavior only, we expect the accurate value to be slightly larger (as the second behavior is slower).\\

Then, the large mass regime begins, in which the differential equation is well approximated by  
\begin{equation}
    \frac{dM}{dt}=\lambda \left( \frac{9}{16 \pi^2}\rho c \right)^{1/3} M^{2/3}.
    \label{eqdiflarmem}
\end{equation}
The solution $M(t)$ is now regular and scales as $M(t)\propto t^3$ for large values of $t$.
Integrated between $t_c$, where this regime begins to hold, and $t$, Eq. (\ref{eqdiflarmem}) leads to:
\begin{equation}
    M(t)=\left(\lambda \left( \frac{1}{48 \pi^2}\rho c \right)^{1/3}(t-t_c) + M_c^{1/3}\right)^3.
    \label{approxigm}
\end{equation}
The divergence at finite time has disappeared. Much more satisfyingly, the mass of the black hole now goes to infinity for $t\rightarrow \infty$. The mass singularity is regularized without any exotic hypothesis.\\

\begin{figure}
    \centering
    \includegraphics[width=.9\linewidth]{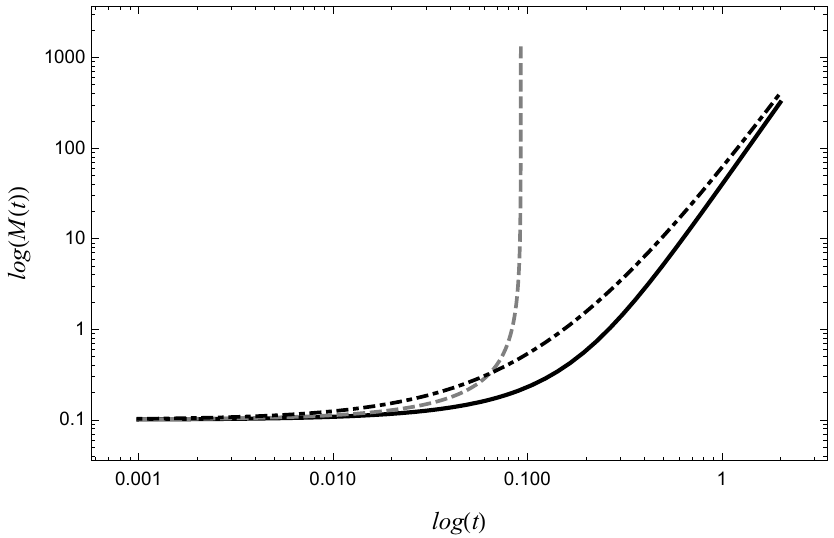}
    \caption{Mass of the black hole as a function of time, in double logarithmic scale and in Planck units -- setting $\rho=1$. The exact solution is in solid black, the low mass approximation is in gray dashed, and the large mass approximation is in black dot-dashed. The divergence of the Schwarzschild solution -- visible on the dashed grey curve -- is not anymore present.}
    \label{fig4}
\end{figure}

Close to $M_c$, where the approximations begin to break down, there is no analytical solution and Eq. (1) has to be solved with the exact mass given by Eq. (\ref{fulleq}). A numerical integration leads to the evolution shown on Fig. (\ref{fig4}). The picture is fully regular.\\

It should be noticed that, in agreement with the explosive nature of the ``small mass" regime, $t_c$ mostly depends upon $M_{init}$ as soon as the initial mass is not too close to the critical mass. It can be rewritten as
\begin{equation}
    t_c = t_{div}-\frac{c^5}{\beta G^2 \rho M_c}.
\end{equation}
As it should, it is smaller than $t_{div}$.

\section{Competition with gravitational waves in a binary system}

Let us now turn to a different question to investigate other consequences of the model. What happens if one considers a binary system of identical black holes emitting gravitational waves and accreting energy from the surrounding medium? The problem was addressed in \cite{Blachier:2023ygh, Barrau:2023ssf} using the Schwarzschild solution. Both phenomena -- the  gravitational waves and the accretion -- play in the same direction and tend do decrease the distance $D$ between black holes. Maybe surprisingly, it was shown in \cite{Blachier:2023ygh} that the mass (and size) divergence of the black holes is {\it always} reached before the coalescence. In the model presented here, there is no more singularity and the dynamics should obviously be very different.\\

Taking into account the power lost by gravitational waves and the momentum conservation, one is led to the following differential equation for the orbital separation:
\begin{equation}
    \dot{D} = -\frac{128}{5} \frac{G^{3}}{c^{5}} \frac{M^{3}}{D^{3}} - 3\frac{\dot{M}}{M}D.
\label{eq:ED_general}
\end{equation}
When the growth rate of $M$ is given by Eq. (\ref{dmdt}), that is in the Schwarzschild case, the solution of Eq. (\ref{eq:ED_general}) is \cite{Blachier:2023ygh}:

\begin{equation}
D(t) = D_{0} \left(1 - \frac{t}{t_{div}} \right)^{3} 
    \left(1 - \frac{t_{div}}{14t_{\mathrm{cc}}} \left[ \left( \frac{t_{div}}{t_{div} -t}\right)^{14} - 1 \right] \right)^{1/4},
\end{equation}
where $D_0$ is the initial separation between the black holes and $t_{\mathrm{cc}}$ is the time of coalescence of the binary system \textit{if} the two BHs were of constant mass $M_0$, {\it i.e.} \cite{Maggiore:2007ulw}:
\begin{equation}
    t_{\mathrm{cc}} = \frac{5}{512} \frac{c^{5} R_{0}^{4}}{G^{3}M_{0}^{3}}.
\label{eq:def_tcc}
\end{equation}\\

If one now assumes that the mass evolution is given by Eq. (\ref{approxigm}), that is by the model proposed in this article -- in the regime where it differs from the usual view --, Eq. (\ref{eq:ED_general}) remains of the Bernoulli
type and is still solvable analytically. The solution is then given by (with the initial mass assumed to be $M_0$ at $t=0$)
\begin{equation}
D(t) = D_{0} \left(\frac{\tau}{t + \tau} \right)^{9}
    \left(1 - \frac{1}{46} \frac{\tau}{t_{\mathrm{cc}}} \left[ \left( 1 + \frac{t}{\tau} \right)^{46} - 1 \right] \right)^{1/4},
\end{equation}
with 
\begin{equation}
\tau=\frac{3}{\lambda}\left(\frac{16\pi^2M_0}{9\rho c}\right)^{1/3}.
\end{equation}
The evolution is now free of any pathology. The system merges without encountering a singularity.

\section{Metric and thermodynamical properties}

If one were to write the metric naturally associated with this model, it would read 
\begin{equation}
    ds^2=\left(1-\frac{R_H}{r}\right)c^2dt^2-\frac{dr^2}{\left(1-\frac{R_H}{r}\right)}-r^2d\Omega^2,
\end{equation}
with, once again, $R_H$ given by Eq. (\ref{fulleq}). This also allows to define a Hawking temperature \cite{Hawking:1975vcx}, 
\begin{equation}
T=\frac{\hbar\kappa}{2\pi k c},
\end{equation}
where $k$ is the Boltzmann constant, $\hbar$ the reduced Planck constant, and 
\begin{equation}
\kappa=\frac{c^2g_{00}'}{2\sqrt{-g_{00}g_{11}}}|_{r=R_H}
\end{equation}
is the surface gravity. If one focuses on the large mass regime (dense medium), where the model differs from the known behavior, the surface gravity becomes
\begin{equation}
\kappa=\left(\frac{\pi c^4 \rho}{6 M}\right)^{1/3},
\end{equation}
leading to 
\begin{equation}
T=\left(\hbar^3\frac{c\rho}{48\pi^2 k^3M}\right)^{1/3}.
\end{equation}
The temperature is now proportional to $M^{-1/3}$ whereas it is usually proportional to $M^{-1}$. Very massive black holes are therefore hotter, which makes sense as they are smaller than for the Schwarzschild metric. 

In principle it is straightforward to calculate the associated mass variation:
\begin{equation}
\frac{dM}{dt}\propto - R_H^2T^4\propto - M^{-2/3},
\end{equation}
faster than the usual $M^{-2}$ for large masses. This integrates, with $t_{ev}$ the evaporation time, in 
\begin{equation}
M(t)=M_0\left( 1 - \frac{t}{t_{ev}} \right)^{3/5},
\end{equation}
whereas the usual power is 1/3. Instead of scaling as $M^3$, the lifetime of the black hole now scales as $M^{5/3}$, which is much smaller. One should however keep in mind that, unless an unexpected phenomenon prevents the accretion from occurring, the large mass regime considered here is precisely the one for which the flux of energy should be drastically inward-directed.

\section{Quantum gravity and contracting universe speculations}

Let us go one more step ahead in speculations. The Kretschmann scalar, $R_K^2=R^{\mu \nu \rho \lambda}R_{\mu \nu \rho \lambda}$, where $R_{\mu \nu \rho \lambda}$ is the Riemann tensor, is known to be extremely small at the surface of a large black hole. In Planck units (used in this section), $R_K\sim 1/M^2$. This is of course not true anymore for this model and it becomes -- once again focusing on the large mass regime--, to be of the order
\begin{equation}
    R_K\sim \rho.
\end{equation}
Importantly, it does not depend  on the mass of the black hole anymore and is uniquely determined by the density of the surrounding medium. Heavy black holes can now exhibit a very high Kretschmann scalar. This is rooted in the fact that the surface gravity does not  decrease anymore with the mass.\\

Using, as in \cite{Haggard:2016ibp}, the ratio $x=l_{Pl}/l_R$, where $l_{Pl}$ is the Planck length and $l_R$ is the curvature length in the considered region (of order $R_K^{-1/2}$), as an indication of the ``quantumness" of the gravitational field, one is led at the horizon to:
\begin{equation}
x\sim \sqrt{\rho}. 
\end{equation}
Remarkably, one might then expect quantum gravity effect at the horizon of arbitrary large black holes -- which is not at all the case usually, {\it e.g.} for a Schwarzschild stellar mass black hole, $x\sim 10^{-38}$ -- as long as they are surrounded by an extremely dense medium. Interestingly, several independent arguments suggest that quantum gravity corrections might be important at the horizon scale \cite{Haggard:2016ibp,Corda:2023pkv,Corda:2023inq,Corda:2024okr,Vaz:2014rya,articleXX}.\\

Beyond its theoretical interest, this work might even have practical consequences for bouncing cosmological models appearing in a variety of approaches (see, {\it e.g.}, \cite{Brandenberger:2016vhg,Date:2004fj,Barrau:2020nek}). In the contracting phase, before the bounce, the density of the universe inevitably becomes huge and, in most models, reaches Planckian values. In the usual Schwarzschild view, it was shown that the catastrophic divergence of the mass of any black hole is always reached {\it before} the bounce  
\cite{Barrau:2022bfg}, practically ruling out the considered bounce scenario (as soon as black holes are present in the contracting phase -- which is plausible as the cosmological dynamics then helps the formation of black holes). The model we have presented here cures this pathology and suggests that black holes might behave non-singularly at the bounce -- in agreement with more rigorous arguments \cite{Carr:2011hv}. The possibility that quantum effects triggered at their horizon might have left footprints is left for another study.\\

A remark is in order. In the usual case ($R_H\propto M^2$), the mass divergence can be escaped if the density of the surrounding medium {\it decreases} fast enough, {\it e.g.} if radiation is diluted and redshifted in an expanding universe. In particular, if the black hole is surrounded by a thermal bath at temperature $T_B\propto t^{\alpha}$, the divergence can be avoided (depending on the value of the initial mass) when $\alpha < -1/4$. One might wonder whether, the other way around, the divergence cured by our model could be revived if the density of the surrounding medium {\it increases} fast enough ($\alpha >0)$. This is not the case: whatever the considered power law for the evolution of the temperature of the surrounding temperature, the behavior remains regular.

\section{conclusion}

This little work obviously did not aim at giving a final answer to the subtle question of the growth of a black hole in a medium whose density is not negligible when compared to the one of the black hole. This situation is anyway not very relevant for phenomenology but mostly for the formal understanding of the behavior of black holes in this rather extreme case — this is often how progresses are made in theoretical physics. Taking advantage of the fact that the naïve Newtonian vision happens to lead to correct results in the vacuum case, we have extended it with the correct inclusion of the effect of the surrounding medium. Maybe surprisingly, the resulting model is consistent and convincing, at least — we believe — more realistic than the usual view leading to a mass divergence at finite time. Beyond a critical mass, which is explicitly calculated, the horizon radius basically scales and $M^{1/3}$ and the growth rate becomes a power law of time.  Unlike in the usual vision, the mean density of the black hole always remains higher than the one of the surrounding medium. It also implies that quantum gravity effects are possible at the horizon of super-massive black holes as soon as they are surrounded by an extremely dense medium. More importantly, this might save bouncing models from the black hole catastrophe. 

Obviously, this cannot be the full story. Our hope is that this toy-model might help deriving a more rigorous solution. Although of marginal practical importance, this situation appears as an interesting thought experiment.

\section*{acknowledgements}
The authors would like to thank Orphée Barrau, Pierre Jamet, Martin Teusher, and Jame Thezier for helpful discussions

\appendix
\section{Other versions of the initial paradox}
Let us show two other sides of the ``paradox" mentioned at the beginning of this article. First, let us consider once again a ball within an infinite medium of density $\rho$ but, this time, let us assume that this ball has a density $\rho_{ball}$. How does the gravitational field at the surface of the ball varies with $\rho_{ball}$ at fixed external $\rho$? Once again, the Gauss theorem leads to a clear and simple answer: $\mathcal{G}(R_H)$ increases (in absolute value) linearly with $\rho_{ball}$ -- this is obvious as, in this view, the field depends only on the mass contained within the Gauss surface. However for the specific value $\rho_{ball}=\rho$ the entire space is homogeneous and all planes containing the considered point at the surface of the ball are symmetry plans and contain the field. Therefore, for $\rho_{ball}=\rho$, $\mathcal{G}(R_H)=0$. We end up with a very strange picture: the modulus of the field increases linearly with $\rho_{ball}$ except for one point where it is discontinuous and vanishes.\\

Finally, let us now assume that the ball of radius $R_H$ is the core of a finite ``star" of radius $R_{tot}$ and density $\rho$, surrounded by an empty space. As well known -- and obvious by Gauss theorem -- the gravitation field at $R_H$ is non-vanishing, proportional to the mass contained within the ball. It does not depend upon $R_{tot}$. However, for $R_{tot}=\infty$, {\it i.e.} for an infinite homogeneous medium, any plane going through the considered point is a symmetry plane, hence the field vanishes.\\

No mystery here: this simply underlines that the way the limit is taken matters. In an infinite medium full of matter, any observer $O$ sees a locally vanishing field because the limit is taken from her viewpoint. Should she take the limit form another point $O'$, she would obviously conclude that $O'$ is the center of mass, hence the center of attraction. This is true for any ball, of any radius, centered on $O'$. This is our point and the spirit of this work: the limit has to be taken from the point where the observer is located. This leads to the refined horizon position suggested here.

\bibliography{ref.bib}

\end{document}